\documentclass[showpacs,showkeys,preprintnumbers,twocolumn]{revtex4}
\usepackage{dcolumn}
\usepackage[dvips]{epsfig}
\usepackage{latexsym}

\def\beq{\begin{equation}}
\def\eeq{\end{equation}}

\begin{document}

\title{Using FEMLAB for Gravitational problems: numerical simulations for all.}

\author{C. Cherubini and S. Filippi}
\affiliation{Engineering Faculty, University Campus Biomedico, Via E. Longoni 83,  I-00155 Rome, Italy and\\
International Center for Relativistic Astrophysics - I.C.R.A.\\
University of Rome ``La Sapienza'', I-00185 Rome, Italy
}

\date{\today}

\begin{abstract}
We discuss  the possibility to solve Modern Numerical Relativity
problems using finite element methods (FEM). Adopting a "user
friendly" software for handling totally general systems of nonlinear
partial differential equations, FEMLAB, we model and numerically
solve in a short time a Gowdy vacuum spacetime, representing an
inhomogeneous cosmology. Results agree perfectly with existing
simulations in the recent literature based not of FEMs but on finite
differences methods. Possible applications for non relativistic
Astrophysics, General Relativity, elementary particle physics and
more general theories of gravitation like EMDA and branes are
discussed.

\end{abstract}



\maketitle
\section{Introduction}
General Relativity (GR) deals with dynamical deformations of space
and time, massively using, as it happens for continuum mechanics,
tensor calculus. Mathematically, General Relativity is described by
a very large number of coupled strongly nonlinear PDEs. The
extraordinary developments of Numerical Relativity of the last ten
years, due to rapid developments of large scale computational
resources, have shown that some tools developed in the past by
mechanical engineers, and in particular Finite Elements techniques,
can be fruitfully adopted in this field too. The spontaneous
formation of event horizons, Cauchy horizons and infinite curvature
singularities, still poses serious problems both at theoretical and
numerical level. For this reason Numerical Relativity is today a
branch of theoretical physics which is growing up in complexity,
requiring knowledge of both high level GR theoretical background as
well as purely numerical one. Numerics in particular still
represents a terrible obstacle for those scientists which want to
enter such a stimulating new area. In this article we show that
FEMLAB \footnote{In its last release, FEMLAB has been renamed as
COMSOL Multiphysics $3.2$. See http://www.comsol.com/ for details.},
a user friendly software developed for solving general systems of
nonlinear partial differential equations from zero (ODEs) up to
three dimensions, with or without time dependence,
 in general as well as in weak form, can be usefully applied to
gravitational physics problems. In \cite{CH1,CH2} FEMLAB was used
for studying the scattering on a sonic black hole, applying standard
procedures developed for analyzing wave fields around Kerr black
hole, i.e. a $3+1$ constrained evolution scheme and excision in
horizon penetrating coordinates\cite{TEUK}. Although the problem was
linear, the presence of an event horizon required an extremely
accurate use of FEMLAB due to necessity of very high resolution and
precision in order to numerically keep all the perturbations
confined inside the black hole once they have crossed its event
horizon and avoid constraint violations. Here instead, we attack an
exact nonlinear problem, i.e. we study a $1+1$ set of Einstein's
vacuum field equations describing a Gowdy inhomogenous cosmology,
showing how to model in less then ten minutes the problem with
FEMLAB. Other ten minutes of running on a standard laptop pc are
required in order to be ready to compare the results of the
simulation with existing literature based on standard and advanced
finite differences methods. It appears clear that FEMLAB is able to
solve perfectly the problem capturing the fine structure (the
spikes) characteristic of this class of universes. This application
suggests that FEMLAB can become for sure a necessary tool for those
scientists and mathematicians working in the field of full GR
analysis as well as in partial differential equations present in non
relativistic astrophysics, elementary particle physics, statistical
mechanics as well as extended theories of gravitation (Brans-Dicke,
branes, EMDA, etc...).

\section{ Field Equations}
In this section we introduce the mathematical framework for our
FEMLAB simulations. We follow reference \cite{Garfinkle} (and
references therein) for derivation of the field equations and
comparison of numerical solutions. The Gowdy spacetime on
${T^3}\times R$ has the form
\begin{equation}
ds^2=\frac{e^{\frac{\lambda}{2}}}{\sqrt{t}}(-dt^2+dx^2)+t[e^P(dy+Qdz)^2+e^{-P}dz^2]
\label{gowdy}
\end{equation}
where $P, Q$ and $\lambda $ are functions of $t$ and $x$. The
$T^3$ spatial topology is imposed by having $0\le x,y,z \le 2 \pi$
and being $P, Q$ and $\lambda$ periodic functions of $x$.   We
introduce the coordinate $\tau \equiv - \ln t$ such that the
singularity is approached as $\tau \to \infty$.  In terms of this
coordinate, vacuum Einstein field equations split into first order
 evolution equations in divergence form (comma denotes ordinary
derivation)
\begin{eqnarray}
&&P_{,\tau }=R\,\,\,,\nonumber\\
&&Q_{,\tau }=S\,\,\,,\nonumber\\
&&{R_{,\tau}}-{e^{-2\tau}}{P_{,xx}}=-{e^{2(P-\tau
)}}{Q_{,x} ^2}+{e^{2P}}{S^2} \,\,\,,\nonumber\\
&&{S_{,\tau}}-{e^{-2\tau}}{Q_{,xx}}=
2{e^{-2\tau}}{P_{,x}}{Q_{,x}}-2{R S}\,\,\,,\label{eqns}
\end{eqnarray}
and constraint equations
\begin{eqnarray}
&&\lambda_{,x }=-2P_{,x}R-2{e^{2P}}Q_{,x}S\,\,,\nonumber\\
&&\lambda_{,\tau
}=-{e^{-2\tau}}P_{,x}^2-{e^{2(P-\tau)}}Q_{,x}^2-{e^{2P}}S^2-R^2\,\,,\label{constr}
\end{eqnarray}
which trivially determine $\lambda$ once $P$ and $Q$ are known. For
this reason we will not be interested in this second set of
equations which does not influence the evolution set once a proper
initial data satisfying the constraints is chosen. In particular, a
meaningful initial  data at $\tau =0$ is $P=0, \, R={v_0}\cos x, \,
Q = \cos x, S=0$ (where $v_0=5$ in our simulations). Periodic
boundary conditions are imposed at $x=0$ and $x=2\pi$, i.e.
$P(\tau,0)=P(\tau,2\pi),\,\tau\ge0$ and similarly for $Q,R$ and $S$,
while \lq\lq cusps\rq\rq  are avoided by imposing zero flux Neumann
boundary conditions at both ends, i.e. $Q_{,x}=P_{,x}=0$ only (in
fact $R_{,x}$ and $S_{,x}=0$ do not appear in the field equations
(\ref{eqns}) for the evolution). This is a delicate point and must
be clarified, because periodicity would require to equate gradients
at both ends without specifying their value (as we did in our case
of zero flux instead). Initial data however is symmetric with
respect to $x=\pi$ and field equations, naively, \lq\lq do not
prefer\rq\rq right or left directions (change $x\leftrightarrow -x$
in equations (\ref{eqns}) and (\ref{constr})), consequently
evolution must be symmetric on both sides. Matching solutions at
both ends, we can avoid cusps only if gradient is zero there (no
flux). Clearly this naive statement can be formalized rigorously by
using the evolution equations (\ref{eqns}) cast in null coordinates
and characteristic variables\cite{Garfinkle}. Due to these
symmetries, the simple zero flux conditions generate a smooth
matching on both ends, so in this sense the initial request of
periodicity for $(P,Q,R,S)$ has become useless. We will continue to
require such a condition only to show how FEMLAB can handle periodic
boundary conditions. We can now start our finite element modelling,
pointing out that an excellent introduction to the involved
mathematical theory of Finite Elements can be found in the first
chapters of ref.\cite{HUGES}

\section{Building the model}
\begin{figure}
\begin{center}
\epsfig{file=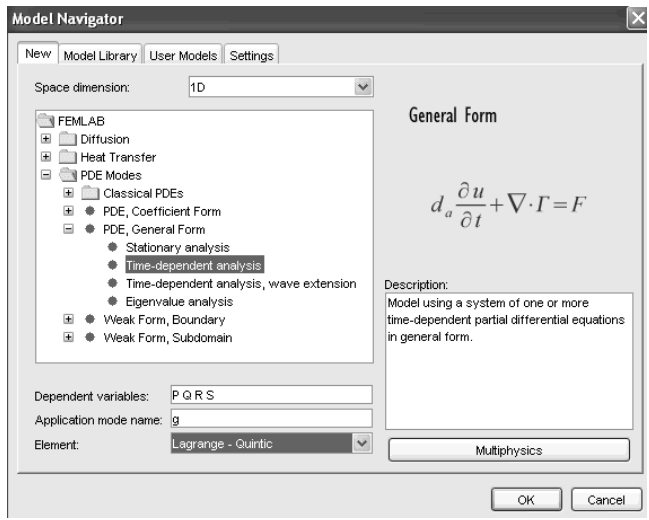,width=8.5cm}
\end{center}
\caption{Choosing the model properties.}
 \label{Fig0}
\end{figure}

\begin{figure}
\begin{center}
\epsfig{file=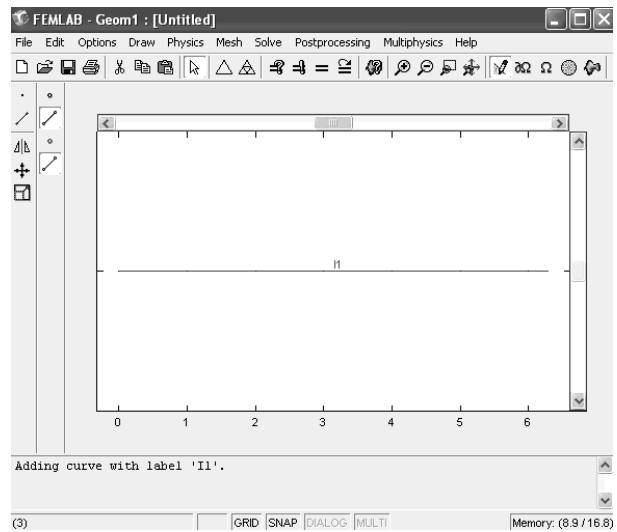,width=8cm}
\end{center}
\caption{Drawing the spatial domain.}
 \label{Fig1}
\end{figure}

\begin{figure}
\begin{center}
\epsfig{file=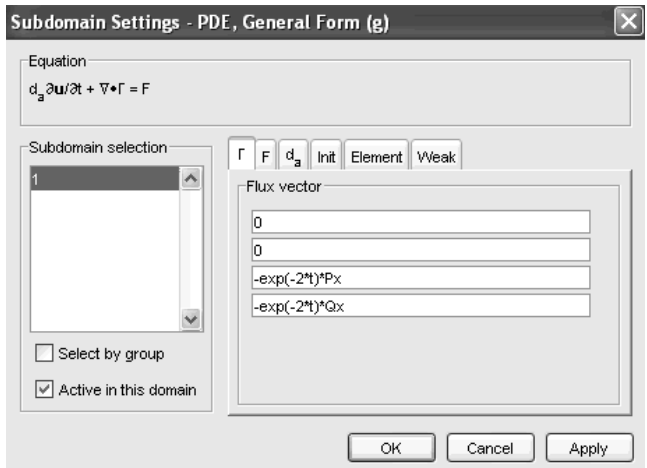,width=8.5cm}
\end{center}
\caption{Entering field equations (first part).}
 \label{Fig2}
\end{figure}

\begin{figure}
\begin{center}
\epsfig{file=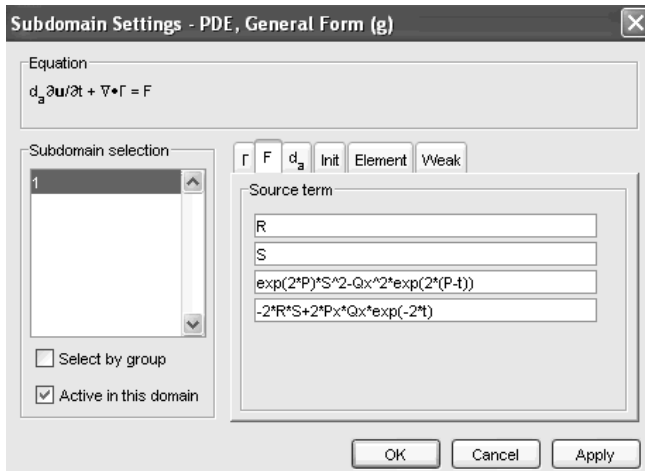,width=8.5cm}
\end{center}
\caption{Entering field equations (second part).}
 \label{Fig3}
\end{figure}

\begin{figure}
\begin{center}
\epsfig{file=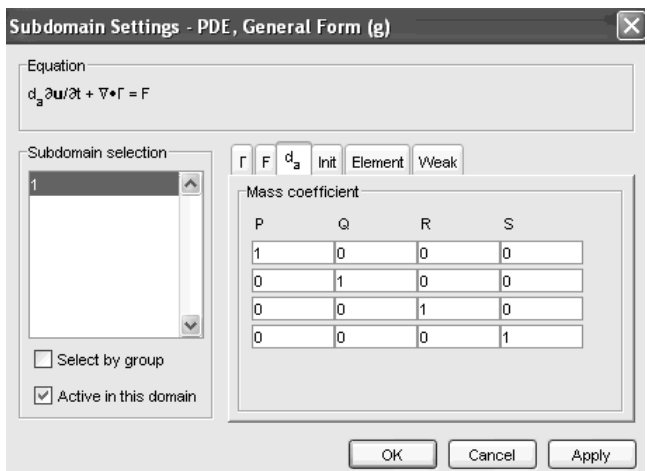,width=8.5cm}
\end{center}
\caption{Entering field equations (third part).}
 \label{Fig4}
\end{figure}

\begin{figure}
\begin{center}
\epsfig{file=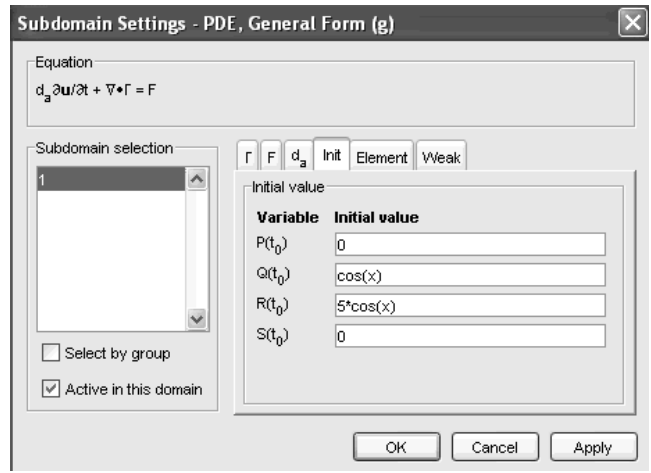,width=8.5cm}
\end{center}
\caption{Entering initial data.}
 \label{Fig5}
\end{figure}

\begin{figure}
\begin{center}
\epsfig{file=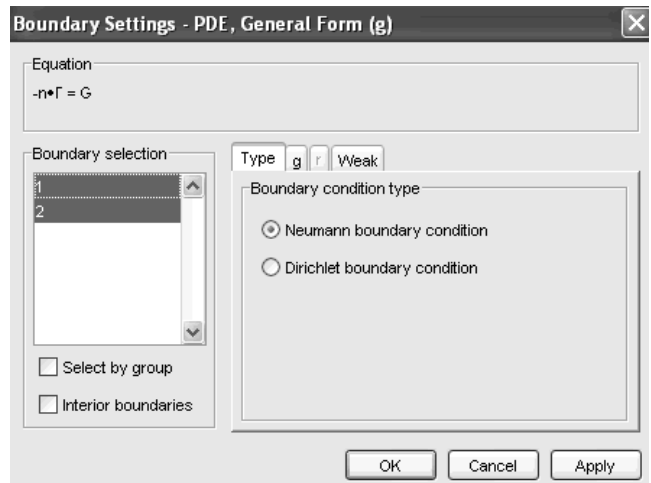,width=8.5cm}
\end{center}
\caption{Selecting zero flux conditions (first part).}
 \label{Fig6}
\end{figure}

\begin{figure}
\begin{center}
\epsfig{file=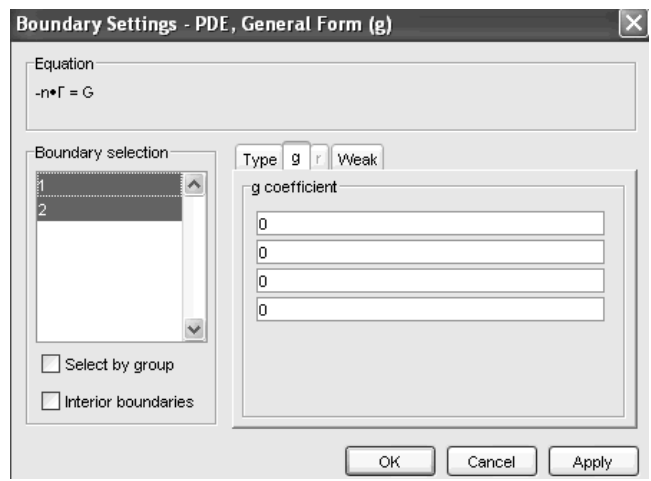,width=8.5cm}
\end{center}
\caption{Selecting zero flux conditions (second part).}
 \label{Fig7}
\end{figure}

\begin{figure}
\begin{center}
\epsfig{file=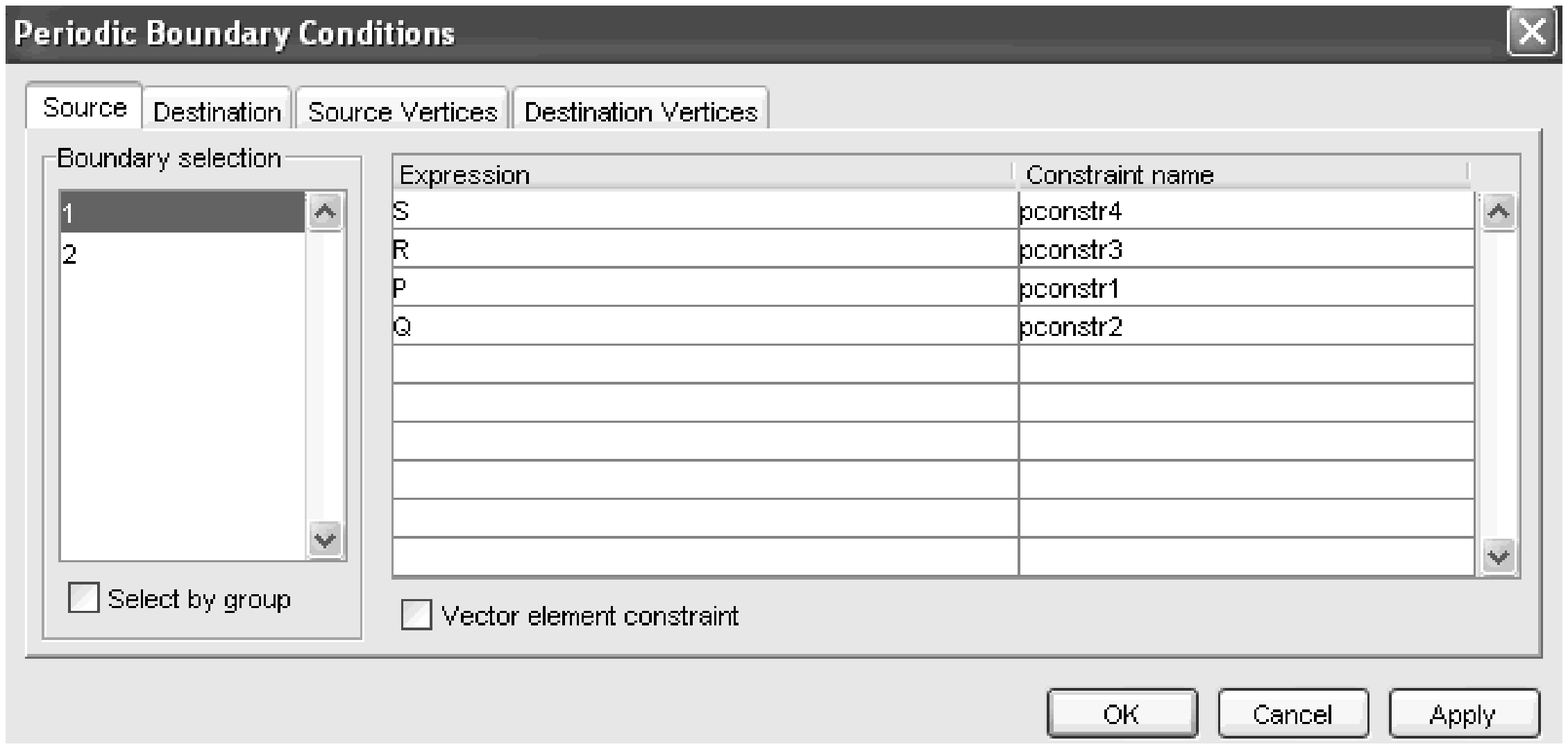,width=8cm}
\end{center}
\caption{Selecting periodic boundary conditions (see text).}
 \label{Fig8}
\end{figure}

\begin{figure}
\begin{center}
\epsfig{file=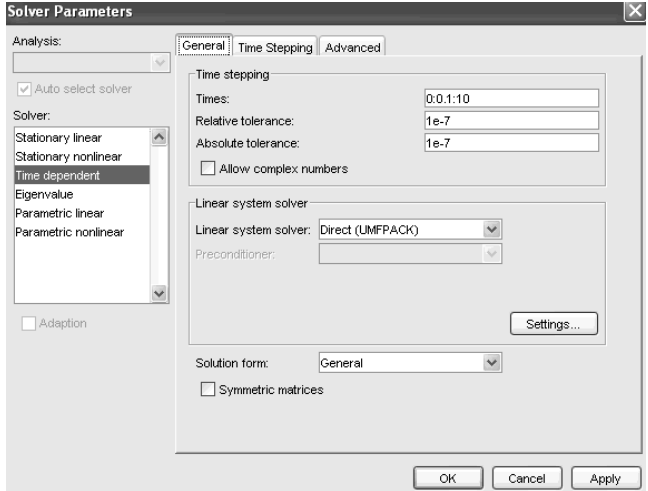,width=8.5cm}
\end{center}
\caption{Selecting numeric properties (first part).}
 \label{Fig9}
\end{figure}

\begin{figure}
\begin{center}
\epsfig{file=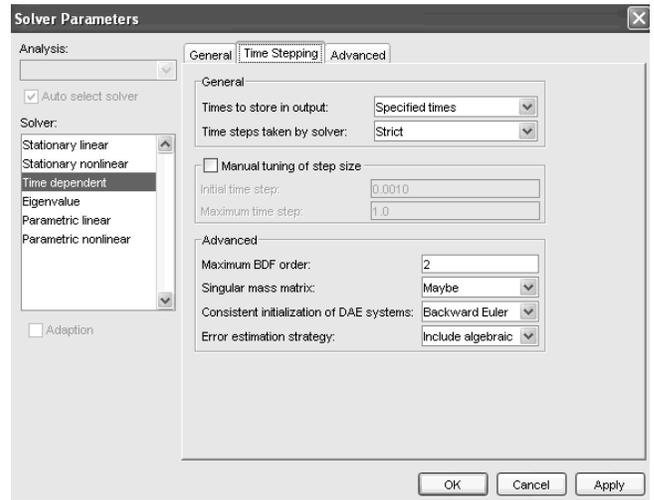,width=8.5cm}
\end{center}
\caption{Selecting numeric properties (second part).}
 \label{Fig10}
\end{figure}

\begin{figure}
\begin{center}
\epsfig{file=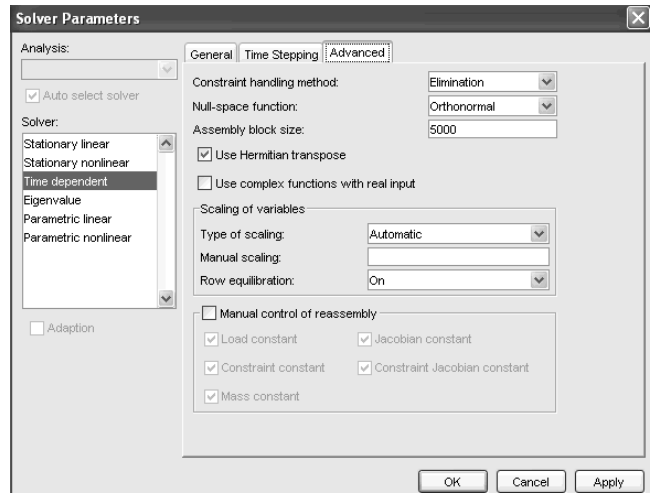,width=8.5cm}
\end{center}
\caption{Selecting numeric properties (third part).}
 \label{Fig11}
\end{figure}

\begin{figure}
\begin{center}
\epsfig{file=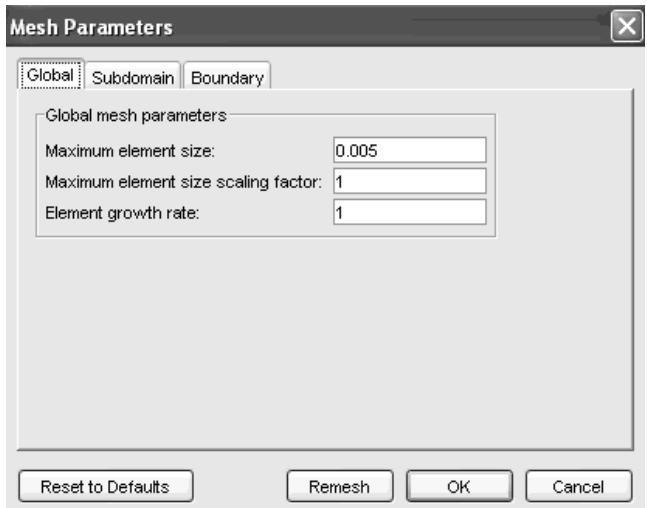,width=8.5cm}
\end{center}
\caption{Selecting meshing properties.}
 \label{Fig12}
\end{figure}

In figure (\ref{Fig0}) we start  selecting the dimensionality of the
problem (1D), the type of problem (time dependent in general form),
the variables $(P,Q,R,S)$ and the element (Lagrange quintic). We
adopted the highest nonlinear element in order to get the best
results, although a quadratic Lagrange element leads practically the
same results (with smaller computational time).
In figure (\ref{Fig1}) we draw the domain, i.e. a line starting at
$x=0$ and ending at $x=2\pi$.
In figures (\ref{Fig2})-(\ref{Fig4}) we insert the field equations
in divergence form while in figure (\ref{Fig5}) we enter the
initial data.
In these FEMLAB modeling pictures as well as in the forthcoming text
and figures, time $t$ has \underline{always} to be intended as
scaled time $\tau$ of equations (\ref{eqns}). In figures
(\ref{Fig6}) and (\ref{Fig7}) instead we chose zero flux Neumann
boundary conditions at both ends of the domain, while in figure
(\ref{Fig8}) one enters the periodic boundary conditions (which as
described before, are automatically implied by zero flux ones). We
show one figure only because the entire procedure would take several
slides due to the fact that one has to make the value of the various
field coincide at both ends (vertices). The procedure is very
simple, although takes few minutes and is well explained in the
online manual with several examples so we refer to it for this part
of the model.
In figures (\ref{Fig9})-(\ref{Fig11}) we select the numerics of the
problem. A direct solver (UMFPACK) is chosen, time integration
starts at $t=0$ and ends at $t=10$, saving on hard disk the
simulation results at every $0.1$ time step. Other options are
chosen in order to have the most precise settings for the
simulation. Time stepping is automatically chosen by the solver
which refines it when necessary (although one can always set it up
manually). Concerning numerical accuracy, we have selected relative
and absolute tolerance thresholds of $10^{-7}$. This choice will be
sufficient to get high quality results.
In figure (\ref{Fig12}) we select the meshing of the domain, which
is uniform and has a spacing of $\Delta x=0.005$. While FEMLAB can
handle adaptive mesh refinement, we prefer to use such a very fine
constantly spaced meshing here.
\section{Results}
\begin{figure}
\begin{center}
\epsfig{file=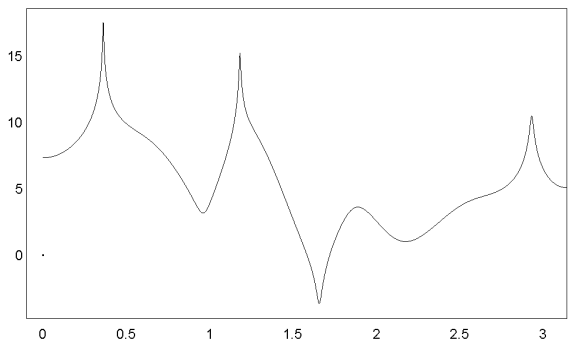,width=8cm}
\end{center}
\caption{Plot of $P$ at $t=10$ (read $\tau$) for $0\le x\le\pi$.}
 \label{Fig13}
\end{figure}

\begin{figure}
\begin{center}
\epsfig{file=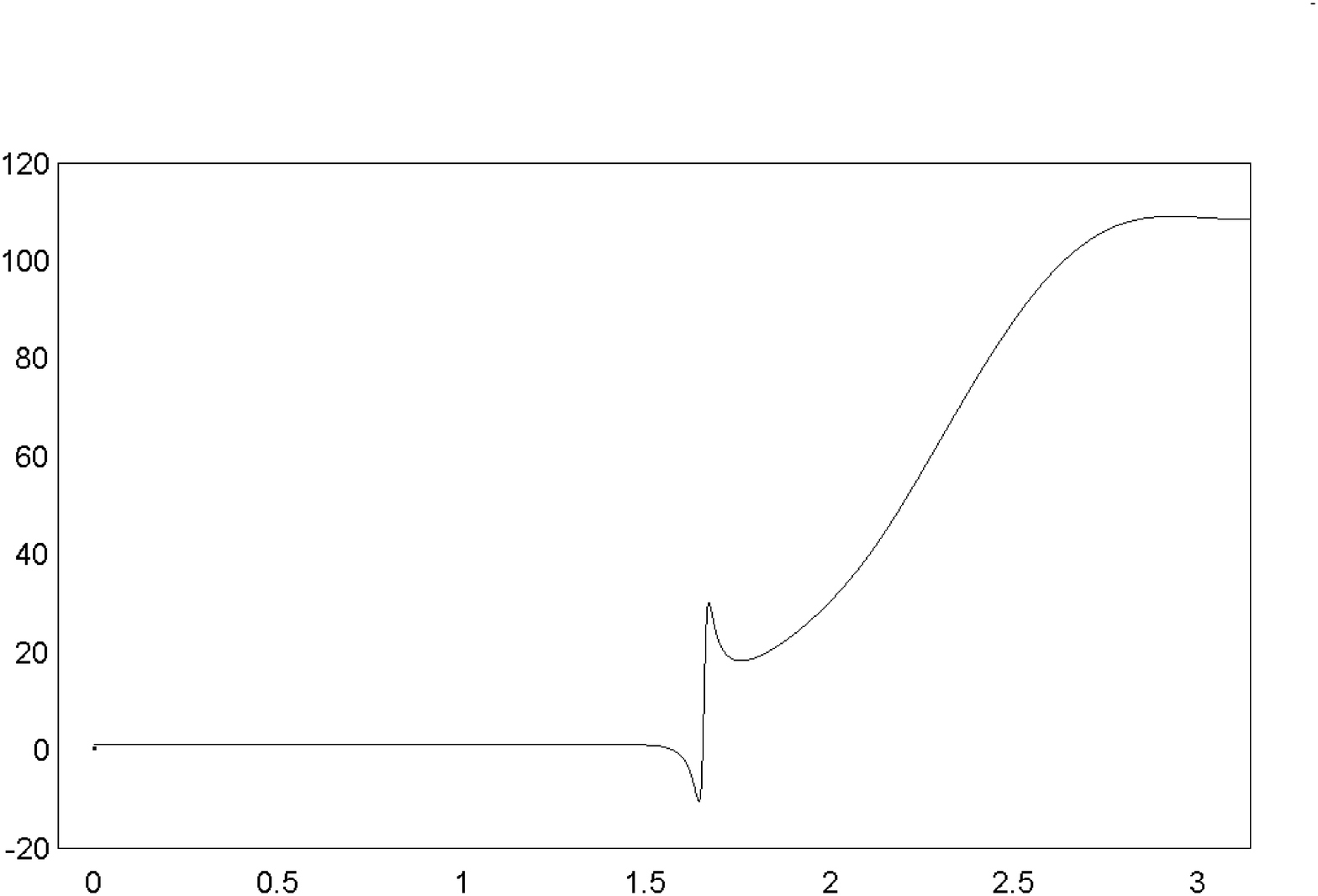,width=8cm}
\end{center}
\caption{Plot of $Q$ at $t=10$ (read $\tau$) for $0\le x\le\pi$.}
 \label{Fig14}
\end{figure}

\begin{figure}
\begin{center}
\epsfig{file=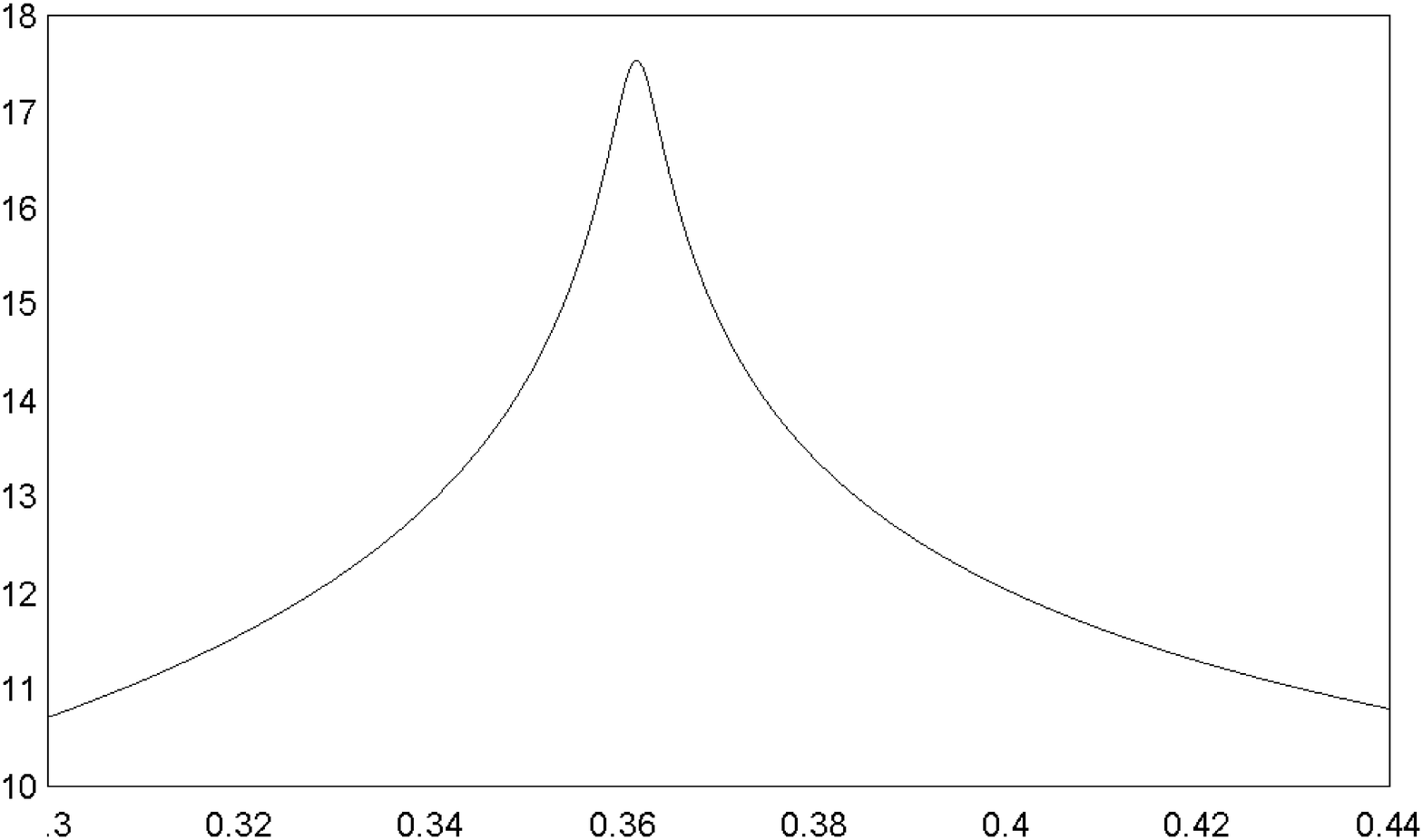,width=8cm}
\end{center}
\caption{Detail of the plot for $P$ at $t=10$ (read $\tau$) around
$x=0.36$.}
 \label{Fig15}
\end{figure}

\begin{figure}
\begin{center}
\epsfig{file=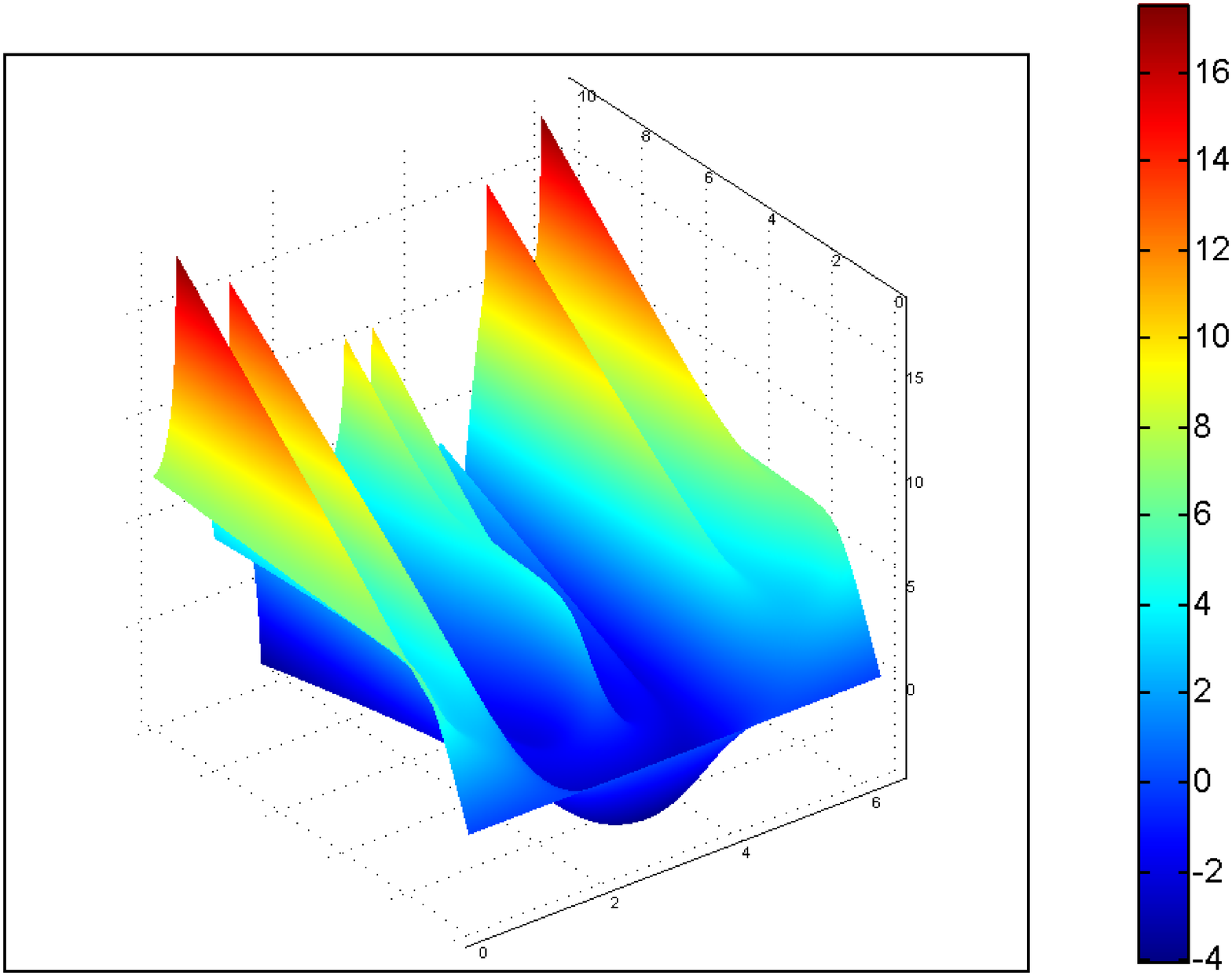,width=8cm}
\end{center}
\caption{Time evolution of $P$ in the interval $0\le x\le2\pi$.}
 \label{Fig16}
\end{figure}

\begin{figure}
\begin{center}
\epsfig{file=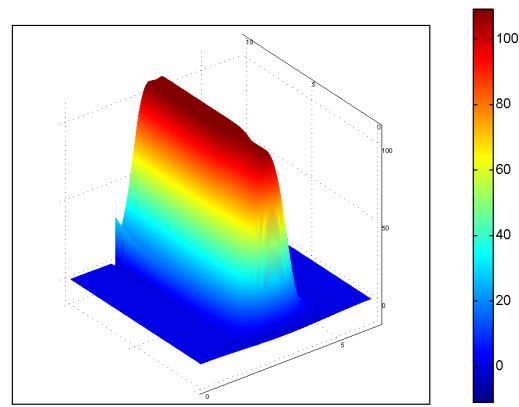,width=8cm}
\end{center}
\caption{Time evolution of $Q$ in the interval $0\le x\le2\pi$.}
 \label{Fig17}
\end{figure}

As anticipated before, simulation takes approximately ten minutes to
run on a Pentium $4$ running at $1.7$GHz. In figure (\ref{Fig13}) we
show $P$ at $t=10$, in the interval $0\le x\le\pi$, while figure
(\ref{Fig14}) is a plot of $Q$ with same conditions. Figure
(\ref{Fig15})  shows a detail (a spike) of figure (\ref{Fig13}).
These three figures are exactly coincident with figures $1,2$, and
$3$ of reference  \cite{Garfinkle}, i.e. FEMLAB produced the same
results of finite differences simulations using finite elements
instead. In figures (\ref{Fig16}) and (\ref{Fig17}) we show the time
evolution in the entire spatial domain of $P$ and $Q$ respectively.
This visualization clearly shows the generation and subsequent
development of these Gowdy spikes. The initial data for $Q$, i.e.
$\cos x$ is not evident in figure (\ref{Fig17}) due to the very
large change of scale, which has passed from $1$ to around $100$ in
ten time units.
\section{Discussion}
The simulation presented in this article and the comparison with
existing literature suggests that FEMLAB can become for sure a
necessary tool for those scientists and mathematicians working in
the field of GR analysis, as in the past years Maple with its tensor
packages did. Due to its possibility to handle general PDEs,
problems in astrophysics (equilibrium configurations, MHD, ...) and
in more general theories of gravitation like EMDA and branes can be
solved. Being FEMLAB totally integrated in MATLAB, and having MAPLE
a direct export to it, the process of symbolic derivation of field
equation, numerical solution and export for deeper analysis ( for
studying the solution in the frequency domain in order to extract
quasi-normal modes, for example) and comparison with experimental
data results straightforward. Moreover, being FEMLAB designed for
$64$bit platforms, large amounts of memory (typically a ten of Gb)
can be addressed for simulations in 2D and 3D, although for the
latter case, having in mind General Relativity problems like
strongly distorted black holes and neutron stars, a parallelized
version of the software would certainly be welcome. Even for those
working in Numerical Relativity, the use of FEMLAB allows them to
test an idea in a simplified version in few hours, and if the
obtained results appear to be promising, write down then an
optimized code for their large scale supercomputers.

\section{Acknowledgments}
CC would like to thank the organizers of the 9th Italian-Korean
Symposium on Relativistic Astrophysics, and Prof H.W. Lee in
particular, for their continuous and efficient assistance during the
entire conference in Korea. All trademarks and registered trademarks
are the property of their respective owners.

\end{document}